\documentclass[english]{llncs}
\usepackage[T1]{fontenc}
\usepackage[latin9]{inputenc}
\usepackage{url}
\usepackage{amsmath}
\usepackage{amssymb}
\usepackage{graphicx}
\usepackage{esint}
\usepackage{babel}
\begin{document}

\title{Diffeomorphic Medial Modeling}

\author{Paul A. Yushkevich$^{1}$, Ahmed Aly$^{1}$, Jiancong Wang$^{1}$,
Long Xie$^{1}$, Robert C. Gorman$^{2}$, Laurent
Younes$^{3}$, Alison Pouch$^{1}$}

\institute{$^{1}$Department of Radiology, University of Pennsylvania Perelman
School of Medicine. \\
$^{2}$Department of Surgery, University of Pennsylvania Perelman
School of Medicine. \\
$^{3}$Department of Applied Mathematics and Statistics, Johns Hopkins
University}
\maketitle
\begin{abstract}
Deformable shape modeling approaches that describe objects in terms
of their \emph{medial axis} geometry (e.g., \emph{m-reps} \cite{pizer03})
yield rich geometrical features that can be useful for analyzing the
shape of sheet-like biological structures, such as the myocardium.
We present a novel shape analysis approach that combines the benefits
of medial shape modeling and diffeomorphometry. Our algorithm is formulated
as a problem of matching shapes using diffeomorphic flows under constraints
that approximately preserve medial axis geometry during deformation.
As the result, correspondence between the medial axes of similar shapes
is maintained. The approach is evaluated in the context of modeling
the shape of the left ventricular wall from 3D echocardiography images.\global\long\def\argmin{\operatornamewithlimits{arg\,min}}
\footnote{This work is supported by NIH grants EB017255 and HL103723.}
\end{abstract}

\section{Introduction}

In medical image analysis, \emph{deformable medial models} (or \emph{medial
representations) }\cite{pizer03,py13ipmi,hong16} are frequently used
to characterize the shape of anatomical structures that have sheet-like
geometry, such as the cerebral cortex, the myocardium, or knee cartilage.
They\emph{ }are deformable models that directly characterize the \emph{medial
axes} of objects. The medial axis (or \emph{skeleton}) of an object
is a set of manifolds formed by all points inside of the object that
are equidistant to two or more points on its boundary \cite{blum78}.
The distance from the medial axis to the boundary describes the local
thickness of objects. For sheet-like anatomical objects, the shape
of the medial axis can be a useful proxy for overall shape, and local
thickness is frequently an important biological measurement.

Medial axes of 3D objects can be easily derived using various skeletonization
approaches (e.g., \cite{naef96}). However, even small perturbations
of the boundary of an object can result in a significant reconfiguration
of the medial axis. This makes it difficult to use medial axes derived
by skeletonization in statistical shape analysis. Deformable medial
modeling techniques such as \emph{m-reps} \cite{pizer03} overcome
this challenge by explicitly controlling the configuration of the
medial axis of the model during deformation (i.e., keeping the number
and connectivity of the surfaces in the medial axis fixed). For many
classes of anatomical objects such models can approximate the shape
of individual objects with good accuracy, while allowing the features
derived from the medial axis to be compared across subjects. Deformable
medial models have been used to perform statistical shape analysis
for various sheet-like brain structures, myocardium and heart valves,
abdominal organs, etc. Medial models have also been used to impose
geometrical constraints on automatic segmentation of sheet-like structures,
e.g., imposing prior knowledge about heart wall thickness during myocardium
segmentation \cite{sun10}. Medial modeling methods include \emph{m-reps}
\cite{pizer03} and \emph{s-reps} \cite{hong16} (which approximate
medial axis surfaces using discrete primitives), \emph{cm-reps }\cite{py06tmi}
(which use splines and subdivision surfaces to model medial axis surfaces),
and \emph{boundary-constrained m-reps }\cite{py13ipmi} (which implicitly
model medial geometry by imposing symmetry constraints on a boundary-based
model). 

This paper addresses a significant limitation of existing medial modeling
approaches: they lack a built-in mechanism to prevent models from
folding or self-intersecting during deformation; and they do not provide
a natural way to extrapolate model transformations to deformations
of the ambient space. In recent years, the field of statistical shape
analysis has embraced an approach based on \emph{flows of diffeomorphisms}.
In this \emph{``diffeomorphometry'' }approach \cite{miller06},
shapes are expressed in terms of smooth invertible transformations
that deform a canonical \emph{template shape} into each shape of interest.
Diffeomorphometry provides a concrete way to impose a metric on the
space of shapes, based on the amount of deformation required to map
one shape into another, and gives rise to rigorous algorithms for
shape interpolation, computation of mean shapes, and other statistics.
Diffeomorphometry also allows transformations between shapes to be
extended naturally to diffeomorphic transformations of the ambient
space, e.g., allowing shape correspondences to be extrapolated to
image correspondences.

We propose an algorithm that combines the benefits of deformable medial
models and diffeomorphometry. Our algorithm is formulated as a problem
of matching shapes using diffeomorphic flows under constraints that
approximately preserve medial axis geometry during deformation. Our
approach leverages a general framework for incorporating geometric
constraints into diffeomorphic shape matching developed by Arguillère
et al. \cite{arguillere15}. To our knowledge, this is the first paper
to integrate diffeomorphometry and deformable medial modeling. Our
experiments focus on demonstrating the feasibility of this method
in the context of matching synthetic and real-world anatomical shapes.

\section{Materials and Methods}

In this section, we first briefly introduce \emph{flows of diffeomorphisms}
for shape analysis. We then review the basic concepts of \emph{medial
geometry} and define a special class of \emph{diffeomorphic flows
that preserve medial geometry}. We then frame the problem of diffeomorphic
medial modeling for point-set representations of 3D shapes and implement
it using the general \emph{constrained diffeomorphometry} framework
developed by Arguillère et al. \cite{arguillere15}.

\subsection{Flows of Diffeomorphisms for Shape Analysis}

In diffeomorphometry, collections of shapes (e.g., hippocampi of different
individuals) are represented as diffeomorphic transformations of some
template shape ${\cal O}$. Following \cite{arnold66}, diffeomorphic
transformations $\{\phi:\mathbb{R}^{n}\rightarrow\mathbb{R}^{n}\}$
form a group under composition. \emph{Flows of diffeomorphisms} are
generated by the ordinary differential equation:
\begin{equation}
\frac{\partial\phi(x,t)}{\partial t}=v(\phi(x,t),t)\,,\qquad\phi(x,0)=x\,,\qquad t\in[0,1],\label{eq:diff_pde}
\end{equation}
where $v:\mathbb{R}^{n}\times[0,1]\rightarrow\mathbb{R}^{n}$ is a
time-varying vector field, called the \emph{velocity} \emph{field},
that satisfies certain smoothness constraints \cite{miller06}. Let
us write $\phi_{t}$ as shorthand for $\phi(\cdot,t)$ and $v_{t}$
for $v(\cdot,t)$. Applying the diffeomorphism $\phi_{t}$ to ${\cal O}$
yields a new shape $\phi_{t}{\cal O}=\{x\in\mathbb{R}^{n}:\phi_{t}^{-1}(x)\in{\cal O}\}$.
When comparing shapes, computing shape statistics, or performing shape-based
segmentation, a common need is to find a deformation of ${\cal O}$
that matches some target shape ${\cal O}_{\mathrm{trg}}$ (or some
target image $I_{\mathrm{trg}}$) at time $t=1,$ while applying the
least total deformation to the space $\mathbb{R}^{n}$. This is formulated
by treating vector fields $v_{t}$ as elements of a reproducing kernel
Hilbert space ${\cal H}$ with kernel $K$, and defining the \emph{kinetic
energy} of a diffeomorphic flow generated by $v$ as $E_{\mathrm{kin}}[v]=\frac{1}{2}\intop_{0}^{1}\|v_{t}\|_{{\cal H}}^{2}\,\mathrm{d}t\,$,
where $\|\cdot\|_{{\cal H}}$ is the norm in ${\cal H}$. Since pairs
of shapes might not have exact correspondence, the shape matching
problem is often relaxed to that of approximate shape matching by
adding a data attachment term $g(\phi_{1}{\cal O})$ that measures
goodness of fit between the deformed template $\phi_{1}{\cal O}$
and ${\cal O}_{\mathrm{trg}}$ (or $I_{\mathrm{trg}}$) to the energy
functional:

\begin{equation}
E[v]=\alpha\,g(\phi_{1}{\cal O})+E_{\mathrm{kin}}[v].\label{eq:lddmm}
\end{equation}

Diffeomorphic flows that minimize $E[v]$ are geodesics in the group
of diffeomorphisms. This gives rise to a powerful machinery for statistical
shape analysis, in which distances between shapes ${\cal O}$ and
${\cal O}_{\mathrm{trg}}$ are defined as the kinetic energy of the
diffeomorphic flow that minimizes (\ref{eq:lddmm}) \cite{miller06}.
Using this machinery, well-posed algorithms for statistical shape
analysis have been developed \cite{miller06}.

\subsection{Medial Axis Geometry}

The medial axis is the locus formed by the centers of all maximal
inscribed balls in an object \cite{blum78}. Formally, consider a
closed object ${\cal O\subset\mathbb{R}}^{3}$ with no holes and a
smooth boundary ${\cal B_{{\cal O}}}$. Let $B_{x,r}=\{y\in\mathbb{R}^{3}:|y-x|\le r\}$
denote a ball with center $x$ and radius $r$. $B_{x,r}$ is called
a \emph{maximal inscribed ball (MIB)} \emph{in }${\cal O}$ if $B_{x,r}\subset{\cal O}$
and there exists no larger ball $B_{x',r'}\subset{\cal O}$ that contains
$B_{x,r}$. The \emph{medial axis} of ${\cal O}$ is the set of all
tuples $(x,r)\in\mathbb{R}^{3}\times\mathbb{R}^{+}$ such that $B_{x,r}$
is an MIB in ${\cal O}.$ We will call the $x$-component of the medial
axis (i.e., set of all MIB centers) the \emph{medial scaffold} of
${\cal O}$, and denote it ${\cal M_{O}}$, and we will call the $r$-component
the \emph{radial scalar field}, denoted ${\cal R_{O}}$. Generically\footnote{A property of ${\cal O}$ is considered ``generic'' if it is invariant
to small smooth perturbations of ${\cal O}$. For example, the centers
of the MIBs in a perfect cylinder form a line, but a small perturbation
breaks this perfect symmetry. }, ${\cal M_{O}}$ consists of a set of surfaces, called \emph{medial
surfaces}. These medial surfaces are bounded by curves that include
curve segments that are shared by multiple medial surfaces, called
\emph{seam curves}, and curve segments that belong to just one medial
surface, called \emph{free edges}. As shown in \cite{giblin03a},
MIBs centered on the interior of medial surfaces are tangent to ${\cal B_{{\cal O}}}$
at two points. MIBs centered on seam curves are tangent to ${\cal {\cal B_{{\cal O}}}}$
at three points, and MIBs centered along free edges are tangent to
${\cal B_{{\cal O}}}$ at just one point. Each point in ${\cal B_{{\cal O}}}$
belongs to exactly one MIB.

\subsection{Medial Structure Preservation under Diffeomorphic Flow}

A flow $\phi$ acting on an object ${\cal O}$ is called \emph{medial-preserving}
if $\phi_{t}{\cal M}_{{\cal O}}={\cal M}_{\phi_{t}{\cal O}}$ for
all $t\in[0,1]$, i.e., the medial scaffold of the transformed object
coincides with the transformed medial scaffold of the original object.
Medial preserving flows maintain point-wise correspondences on the
medial scaffold as the object deforms, allowing statistical analysis
of properties derived from the medial axis, such as thickness. By
contrast, arbitrary diffeomorphic flows that are not medially preserving
may cause the medial scaffold to change structure, e.g., by adding
new surfaces or reconfiguring their geometry \cite{giblin03a}. \textbf{The
aim of this paper is to perform approximate shape matching using medial-preserving
flows.}

Let the tuple of points $\{b_{1},\ldots,b_{n};x\}$ be called a \emph{medial
tuple}\textbf{ }in ${\cal O}$ if there exists an MIB centered on
$x$ that is tangent to ${\cal B}_{{\cal O}}$ at points $b_{1},\ldots,b_{n}$.
Let $\phi$ be a flow of diffeomorphisms that satisfies the following
property: for every medial tuple $Z=\{b_{1},\ldots,b_{n};x\}$ in
${\cal O}$ and every time $t\in[0,1]$, the transformed tuple at
time $t$, denoted $\phi_{t}Z=\{\phi_{t}(b_{1}),\ldots,\phi_{t}(b_{n});\phi_{t}(x)\}$,
is also a medial tuple in $\phi_{t}{\cal O}$. Then it can be shown
(proof omitted due to limited space) that $\phi$ is medially preserving.
Let us denote $\phi_{t}Z$ as $Z^{t}=\{b_{1}^{t},\ldots,b_{n}^{t};x^{t}\}$.
To check if the tuple $Z^{t}$ (with $n>1$) is a medial tuple in
$\phi_{t}{\cal O}$ it suffices to check three conditions:

\begin{alignat}{2}
(x^{t}-b_{i}^{t}) & \perp T_{b_{i}^{t}}\qquad & \forall i\in[1,n]\label{eq:cond3-1}\\
|x^{t}-b_{i}^{t}| & =|x^{t}-b_{1}^{t}|\qquad & \forall i\in[2,n]\label{eq:cond3-2}\\
|x^{t}-b'| & \ge|x^{t}-b_{1}^{t}|\qquad & \forall b'\in\phi_{t}{\cal B}_{{\cal O}}\,,\label{eq:cond3-3}
\end{alignat}
where $T_{b_{i}^{t}}$ denotes the tangent plane to the boundary surface
\textbf{$\phi_{t}{\cal B}_{{\cal O}}$} at the point $b_{i}^{t}$.
Conditions (\ref{eq:cond3-1}) and (\ref{eq:cond3-2}) ensure that
$x^{t}$ is the center of a ball that is tangent to $\phi_{t}{\cal B}_{{\cal O}}$
at points $b_{1},\ldots,b_{n}$, and condition (\ref{eq:cond3-3})
ensures that this ball is inscribed in $\phi_{t}{\cal O}$, and thus
an MIB in $\phi_{t}{\cal O}$. Note that conditions (\ref{eq:cond3-1})
and (\ref{eq:cond3-2}) only involve local geometry of $\phi_{t}{\cal B}_{{\cal O}}$
around the tuple, while condition (\ref{eq:cond3-3}) requires the
knowledge of the entire $\phi_{t}{\cal B}_{{\cal O}}$. In practice,
it appears to be sufficient to only enforce the first two conditions,
and we conjecture that diffeomorphic flows that satisfy only (\ref{eq:cond3-1})
and (\ref{eq:cond3-2}) for all $Z^{t}$ with $n>1$ are medially
preserving.

\subsection{Optimal Control Framework for Diffeomorphometry\label{subsec:Optimal-Control-Framework}}

Above, we defined a set of local geometric constraints that we want
flows of diffeomorphisms to satisfy. We now briefly summarize a framework
for constrained diffeomorphic shape matching developed by Arguillère
et al. in \cite{arguillere15}. We restrict our attention to shapes
represented by point sets, which is relevant for the implementation
of medial constraints in this paper.

Let $q^{0}=\{q_{1}^{0},\ldots,q_{k}^{0}\}$ be a finite set of points
in $\mathbb{R}^{n}$, e.g., a set of landmarks sampled on the boundary
(and perhaps on the interior) of the template shape ${\cal O}$. Let
$q^{t}$ denote the positions of these points at time $t$, i.e.,
$q^{t}=\phi_{t}q^{0}=\{\phi_{t}(q_{1}^{0}),\ldots,\phi_{t}(q_{k}^{0})\}$.
Let us express the data attachment term $g(\phi_{1}{\cal O})$ purely
in terms of these landmarks, i.e., $g(\phi_{1}{\cal O})=g(q^{1})$.
The matching of ${\cal O}$ to the target shape/image then takes the
form of minimizing the energy
\begin{equation}
E[v]=\alpha\,g(q^{1})+\frac{1}{2}\intop_{0}^{1}\|v_{t}\|_{{\cal H}}^{2}\,\mathrm{d}t\,,\label{eq:lddmm-1}
\end{equation}
where $\phi_{t}$ is related to $v$ via (\ref{eq:diff_pde}). This
is a kind of optimal control problem, in which $v$ is a ``control'',
i.e., a function that affects the paths of points $q^{0}$ and guides
them towards their target locations. It is a remarkable fact that
this problem can be restated as a problem involving a much simpler
set of controls associated with each landmark. As proven in \cite{arguillere15},
if $v$ is a minimizer of $E[v]$ in (\ref{eq:lddmm-1}), then $v$
can be interpolated from a set of vector-valued functions $u(t)=\{u_{1}(t):[0,1]\rightarrow\mathbb{R}^{n},\ldots,u_{k}(t):[0,1]\rightarrow\mathbb{R}^{n}\}$
associated with the landmarks. The interpolation is via the kernel
$K$ of the Hilbert space ${\cal H}$. For every $t$,
\begin{equation}
v_{t}(x)=\sum_{j=1}^{k}K(x,q_{j}^{t})\,u^{j}(t)\,.\label{eq:v_interp}
\end{equation}
The function $u_{i}(t)$ is called the \emph{momentum} of the landmark
$q_{i}^{0}$ and can be visualized as a vector placed at each point
along the path traced by $q_{i}^{0}$. In general, the kernel $K(x,y)$
is a matrix-valued function, but in most practical applications, it
is chosen to be of the form $K(x,y)=\eta(|x-y|){\cal I}_{n\times n}$,
where $\eta:\mathbb{R}\rightarrow\mathbb{R}$ is a radial basis kernel.
As in \cite{arguillere15}, we use a Gaussian kernel $\eta(z)=e^{-0.5\,z^{2}/\sigma^{2}}$. 

Given the interpolation formula (\ref{eq:v_interp}), $\|v_{t}\|_{{\cal H}}$
and subsequently $E[v]$ can be expressed entirely in terms of the
momenta $u(t)$:
\begin{equation}
E[u]=\alpha\,g(q^{1})+\frac{1}{2}\intop_{0}^{1}\underbrace{\sum_{i=1}^{k}\sum_{j=1}^{k}u_{i}(t)^{T}K(q_{i}^{t},q_{j}^{t})u_{j}(t)}_{\|v_{t}\|_{{\cal H}}^{2}}\,\mathrm{dt}\,,\label{eq:enrg_u}
\end{equation}
where the landmark positions are updated according to: \vspace{-10bp}

\begin{eqnarray}
\dot{q_{i}^{t}} & = & v_{t}(q_{i}^{t})=\sum_{j=1}^{k}K(q_{i}^{t},q_{j}^{t})\,u^{j}(t)\,.\label{eq:ode_q}
\end{eqnarray}

Diffeomorphometry methods often take advantage of the fact that minimizers
of $E[u]$ are geodesics in the group of diffeomorphisms and thus
are determined by the initial momentum $u(0)$. However, here we are
concerned with minimizing $E[u]$ subject to a set of constraints
$C_{1},\ldots,C_{L}$, each in the form $C_{l}(q^{t})=0$. Arguillère
et al. \cite{arguillere15} propose two strategies for solving such
problems: as an initial momentum problem where geodesics are computed
by projection of $\dot{q}_{i}^{t}$ into the null space of the constraints;
and using an \emph{Augmented Lagrangian (AL)} method in which optimization
is over the complete momentum $u(t)$ and the constraints are added
as penalty terms to $E[u]$. We adopt the AL-based method, which is
easier to implement and is numerically more stable, but converges
slowly \cite{arguillere15}.

In the AL method, minimization of $E[u]$ subject to constraints $\{C_{l}(q^{t})=0\}$
is achieved as a series of unconstrained problems, indexed by $m=1,\ldots,M$.
A set of $L$ Lagrange multiplier functions $\lambda_{l}^{m}:[0,1]\rightarrow\mathbb{R}$
corresponding the constraints is introduced, initialized $\lambda_{l}^{1}(t)=0$
. At iteration $m$, the following energy is minimized:
\begin{equation}
E_{\mathrm{AL}}^{m}[u]=E[u]+\intop_{0}^{1}\sum_{l=1}^{L}\left[\frac{\mu^{m}}{2}C_{l}(q^{t})^{2}-\lambda_{l}^{m}(t)C_{l}(q^{t})\right]\,\mathrm{d}t\,,\label{eq:min_al-1}
\end{equation}
where $\mu^{m}$ is a scalar weight. The Lagrange multipliers and
weights are updated at each iteration as
\[
\lambda_{l}^{m+1}(t)=\lambda_{l}^{m}(t)-\mu^{m}C_{l}(q^{t})\,;\qquad\mu^{m+1}=\mu^{m}\cdot\mathring{\mu},
\]
where $\mathring{\mu}=10$ is a constant scaling factor. Minimization
of (\ref{eq:min_al-1}) only requires computing the gradient of $E_{\mathrm{AL}}^{m}[u]$
with respect to $u(t)$, which involves solving an ODE backwards in
time, as specified in \cite[Remark 24]{arguillere15}. 

\subsection{Medial Constraints for Point Set Diffeomorphometry\label{subsec:discrete}}

In this paper, we adapt the Arguillère et al. \cite{arguillere15}
approach to the problem of matching shapes using medial-preserving
flows of diffeomorphisms. To reduce the problem to point sets, we
relax the constraints (\ref{eq:cond3-1}) and (\ref{eq:cond3-2})
to a finite set of medial tuples $Z_{1}^ {},\ldots,Z_{p}$ in ${\cal O}$,
rather than for the infinite set of all medial tuples in ${\cal O}$.
We also discretize the time dimension. Although by enforcing constraints
(\ref{eq:cond3-1}) and (\ref{eq:cond3-2}) for a discrete set of
tuples and timepoints no longer guarantees the exact preservation
of medial structure by the diffeomorphic flow, in practice, solving
the discrete problem results in near-preservation of medial structure,
i.e., ${\cal {\cal M}}_{\phi_{t}{\cal O}}\simeq\phi_{t}{\cal M}_{{\cal O}}$,
which is sufficient in order to perform statistical shape analysis
using point correspondences between ${\cal M_{O}}$ and $\phi_{t}{\cal M}_{{\cal O}}$.

Given a template object ${\cal O}$, the medial tuples $Z_{1},\ldots,Z_{p}$
in ${\cal O}$ are sampled ``regularly'' from among all medial tuples
in ${\cal O}$. Such sampling is nontrivial, because regular sampling
of the $x$'s on the medial scaffold results in irregular sampling
of $b'$s on the boundary, and vice-versa. Since constructing a template
is a one-off task, we manually sample medial tuples in the template
object ${\cal O}$ using a GUI tool. The tool computes the approximate
medial scaffold ${\cal M_{O}}$ using pruned Voronoi skeletonization
\cite{naef96}. Samples are taken along the seams and free edges in
${\cal M_{O}}$ as well as on the surfaces forming ${\cal M_{O}}$.
Samples are organized into a triangle mesh, which is then inflated
of both sides of each triangle to create a boundary triangle mesh.
Template-building is shown in Fig. \ref{fig:toyevol}(a).

The Arguillère et al. approach \cite{arguillere15} is applied as
follows. We let $q^{0}$ consist of all of the points contained in
the tuples $Z_{1},\ldots,Z_{p}$ . Each tuple $Z_{i}$ indexes $n_{i}$
boundary landmarks and one medial landmark. Let $\mathrm{\mathfrak{B[i,j]}}$
denote the index of the $j-$th boundary landmarks in $Z_{i}$ and
let $\mathfrak{M}[i]$ denote the index of the medial point in $Z_{i}$.
Then, at time $t$, the transformed $i-$th medial tuple is given
by$Z_{i}^{t}=\{q_{\mathfrak{B}[i,1]}^{t},\ldots,q_{\mathfrak{B}[i,n_{i}]}^{t};q_{\mathfrak{M}[i]}^{t}\}\,.$
We formulate a set of constraints that are discrete equivalents of
the constraints (\ref{eq:cond3-1}) and (\ref{eq:cond3-2}). 

Constraint (\ref{eq:cond3-1}) requires the vector $q_{\mathfrak{B}[i,j]}^{t}-q_{\mathfrak{M}[i]}^{t}$
to be orthogonal to $\phi_{t}{\cal B_{O}}$ at the point $q_{\mathfrak{B}[i,j]}^{t}$.
A pair of vectors spanning the tangent plane to $\phi_{t}{\cal B_{O}}$
at $q_{\mathfrak{B}[i,j]}^{t}$ can be approximated as the weighted
sum of the coordinates of adjacent boundary vertices, e.g., using
the Loop tangent scheme \cite[p.71]{zorin00}. We can write these
tangent vectors as $W_{i,j}^{1}q^{t}$ and $W_{i,j}^{2}q^{t}$ where
$W_{i.j}^{1}$ and $W_{i.j}^{2}$ are sparse $3\times k$ matrices.
The discrete equivalents of constraints (\ref{eq:cond3-1}) and (\ref{eq:cond3-2})
are then:\vspace{-10bp}

\begin{equation}
\begin{array}{rcll}
\left(q_{\mathfrak{B}[i,j]}^{t}-q_{\mathfrak{M}[i]}^{t}\right)^{T}W_{i,j}^{\gamma}q^{t} & = & 0 & \quad\forall j\in[1,n_{i}]\,,\gamma=1,2\,,\\
\left|q_{\mathfrak{B}[i,j]}^{t}-q_{\mathfrak{M}[i]}^{t}\right|^{2}-\left|q_{\mathfrak{B}[i,1]}^{t}-q_{\mathfrak{M}[i]}^{t}\right|^{2} & = & 0 & \quad\forall j\in[2,n_{i}]\,.
\end{array}\label{eq:con_quad}
\end{equation}

Following Pizer et al., \cite{pizer03}, we refer to the vectors $q_{\mathfrak{B}[i,j]}^{t}-q_{\mathfrak{M}[i]}^{t}$
as \emph{spokes }(as in spokes on a bicycle wheel). The constraints
state that spokes must be orthogonal to the boundary and that the
lengths of the spokes must be equal. When monitoring AL optimization,
we describe the extent to which the constraints are violated in terms
of two related intuitive metrics:\emph{ spoke-normal deviation}, which
measures the angle between the spoke $q_{\mathfrak{B}[i,j]}^{t}-q_{\mathfrak{M}[i]}^{t}$
and the boundary normal vector at $q_{\mathfrak{B}[i,j]}^{t}$; and
\emph{spoke-spoke mismatch}, which measures the maximum relative error
between the lengths of two spokes in the tuple $Z^{t}$.

\subsection{Data Attachment Term}

Various data attachment terms can be used. When the target locations
$q_{i}^{\text{trg}}$ of the boundary and/or medial landmarks are
given, a simple sum of square distances (SSD) attachment term is used:
$\text{g(\ensuremath{q^{1}})=\ensuremath{\sum_{i\in\varUpsilon}}|\ensuremath{q_{i}^{1}}-\ensuremath{q_{i}^{\text{trg}}|^{2}}\,,}$where
$\varUpsilon$ denotes the subset of landmarks with given target locations.
When fitting a model to a target binary image ${\cal I_{\mathrm{trg}}}$,
we define $g(q^{1})$ as an approximation of the Dice similarity coefficient
(DSC) between the interior of $\phi_{1}{\cal O}$ and the foreground
region of the target binary image. The DSC attachment term leverages
the triangulation of the boundary and medial surfaces created when
medial tuples are sampled. Points in a 3D wedge formed by each medial
triangle and the corresponding boundary triangle are sampled regularly,
and the image ${\cal I_{\mathrm{trg}}}$ is sampled at each sample
point. Each sample point is also assigned a volume element. Integrating
the sampled intensity values times the volume element yields an approximation
of overlap volume, while the volume of $\phi_{t}{\cal O}$ is computed
by just integrating the volume element. The volume of the foreground
region in ${\cal I}_{\mathrm{trg}}$ is known a priori. This allows
DSC and its gradient with respect to $q^{1}$ to be computed.

\subsection{Numerical Implementation}

The AL minimization is implemented in C++ using CPU multi-threading.
ODEs to compute $E_{\mathrm{AL}}^{m}[u]$ and its gradient are solved
numerically using Euler's method with 40 time steps. Unconstrained
minimization (\ref{eq:min_al-1}) at each AL iteration is performed
using a pseudo-Newton method (LBFGS) and is allowed to proceed until
the change in $E_{\mathrm{AL}}[u]$ (or $E_{\mathrm{AL}}[u,\eta,\rho]$)
falls below a tolerance threshold, usually for 1000-4000 iterations.
Constraints, which are a mix of quadratic and linear expressions,
are implemented as sparse matrix-vector multiplication. The scaling
factor $\mathring{\mu}$ in the AL algorithm is set to 10, while the
initial value of $\mu$ is set experimentally, as is the data attachment
term weight, $\alpha$and the standard deviation $\sigma$ of the
Gaussian kernel.

\section{Experiments and Results}

We perform experiments on synthetic and real-world anatomical shapes,
with the focus on demonstrating the feasibility of matching templates
to target shapes using medial-preserving diffeomorphic flows. 

\subsection{Synthetic Shape Example}

\begin{figure}[t]
\includegraphics[width=1\textwidth]{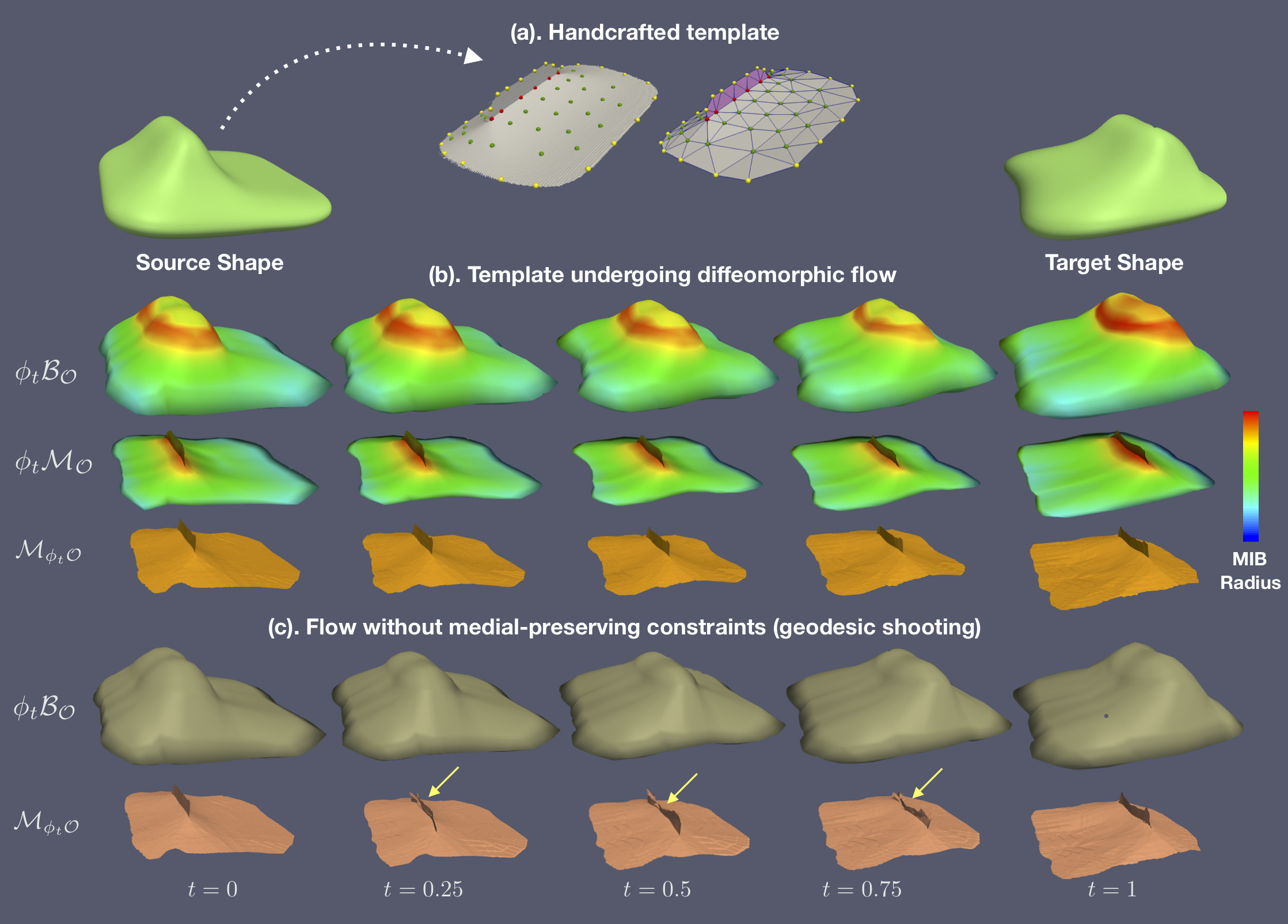}\caption{Template generation and shape matching in the toy example. (a) Manual
template construction by sampling tuples from the skeleton of the
source shape. (b) Medial-preserving diffeomorphic flow between the
template and a target shape. The deforming medial scaffold of the
template ($\phi_{t}{\cal M_{O}})$ is in close agreement with the
Voronoi skeleton of the deforming template (${\cal M}_{\phi_{t}{\cal O}}$),
\emph{which illustrates that medial preservation constraints are working}.
(c) When analogous same shape matching (between ${\cal B}_{{\cal O}}$
and ${\cal B}_{\phi_{1}{\cal O}}$) is performed using point set geodesic
shooting \emph{without medial preservation} \emph{constraints}, the
Voronoi skeleton of the deforming object changes structure and bifurcates
(yellow arrows).\label{fig:toyevol}}
\end{figure}

We use a toy example to show that our method can match shapes with
branching medial axes, a problem that not all medial modeling approaches
can solve (e.g., not \cite{py06tmi}). We created two 3D shapes that
have a medial axis consisting of three surfaces joining along a single
seam curve, as follows. First, we manually painted a binary 3D object
and smoothed it to create a surface labeled ``source shape'' in
Fig. \ref{fig:toyevol}. We then flipped the binary image about the
$x$-axis and performed diffeomorphic deformable image registration
between the original image and the flipped image. The resulting warp
was applied to the source shape, yielding the shape labeled as ``target
shape'' in Fig. \ref{fig:toyevol}. By using deformable registration,
we obtained point-wise correspondences between the two shapes.

To create the initial template medial model, we used a GUI tool as
described in Sec. \ref{subsec:discrete}. The Voronoi skeleton of
the ``source shape'' (after additional smoothing) is shown in Fig.
\ref{fig:toyevol}(a) along with the medial points sampled from the
skeleton and the triangulation of the medial points. The medial mesh
shown in Fig. \ref{fig:toyevol}(a) was inflated and subdivided once
using the Loop scheme \cite{loop90} and then fitted to the ``source
shape'' in Fig. \ref{fig:toyevol} using the deformable medial modeling
method in \cite{py13ipmi}. This yielded an initial template for which
the medial constraints (\ref{eq:con_quad}) are fully satisfied. The
boundary and the medial model of this template are visualized in Fig.
\ref{fig:toyevol}(b) under $t=0$. 

\begin{figure}
\includegraphics[width=1\textwidth]{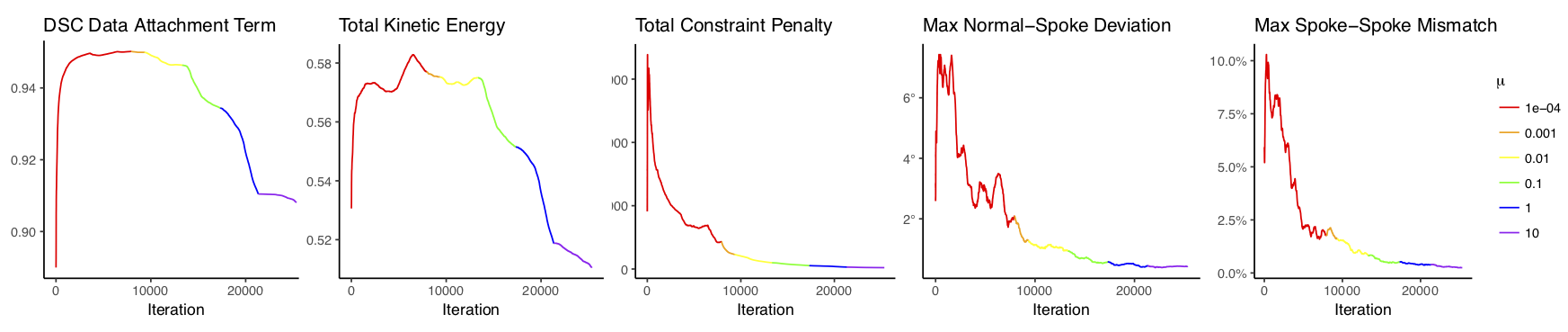}\caption{Data attachment, kinetic energy, and constraints over the course of
AL optimization for synthetic shape matching.\label{fig:toyplots}}
\end{figure}

We then fitted the template to the target shape using our medially-constrained
diffeomorphometry approach. We initialized the fitting using the SSD
data attachment term, with target landmark locations obtained by applying
the image registration warp to the template's boundary landmarks.
This was done using a single AL iteration with $\mu=1$, since the
goal was only to initialize the subsequent DSC-based optimization.
We then performed full AL optimization with the DSC data attachment
term, and with $\mu$ increasing from $0.001$ to $10$ over several
AL iterations. Fig. \ref{fig:toyplots} plots the change in the data
attachment, kinetic energy, total constraint violation, spoke-normal
deviation, and spoke-spoke mismatch over the course of optimization.
As $\mu$ increases ($>0.1$), the data attachment term begins to
retreat from its highest values in order to satisfy the constraints.
However, the resulting practical gain at higher values of $\mu$ is
very small, as the maximal normal-spoke deviation and spoke-spoke
mismatch are already quite small for $\mu=0.1$: about $1^{\circ}$
and $1\%$, respectively, and don't decrease dramatically as $\mu$
increases. The flow visualized in Fig. \ref{fig:toyevol}(b) corresponds
to the solution with $\mu=0.1$, which may be considered the ``sweet
spot'' of the optimization.

The three rows of shapes in Fig. \ref{fig:toyevol}(b) correspond
to the evolving template boundary ($\phi_{t}{\cal B_{O}})$, the evolving
template medial scaffold ($\phi_{t}{\cal M_{O}})$, and the Voronoi
skeleton of the evolving template (an approximation of ${\cal M}_{\phi_{t}{\cal O}}$).
Critically, $\phi_{t}{\cal M_{O}}$ and ${\cal M}_{\phi_{t}{\cal O}}$
are highly consistent, indicating that our\emph{ medial preservation
constraints are working.} By contrast, Fig.\ref{fig:toyevol}(c) shows
that \emph{conventional point set diffeomorphometry does not preserve
medial axis structure}. We performed point set geodesic shooting between
the boundary of our template (${\cal B}_{{\cal O}}$) and the boundary
of the fitted template (${\cal B}_{\phi_{1}{\cal O}}$) and extracted
the Voronoi skeleton of the evolving shape. As pointed out by the
yellow arrows in Fig.\ref{fig:toyevol}(c), the skeleton undergoes
structural change during deformation.

\begin{figure}
\includegraphics[width=1\textwidth]{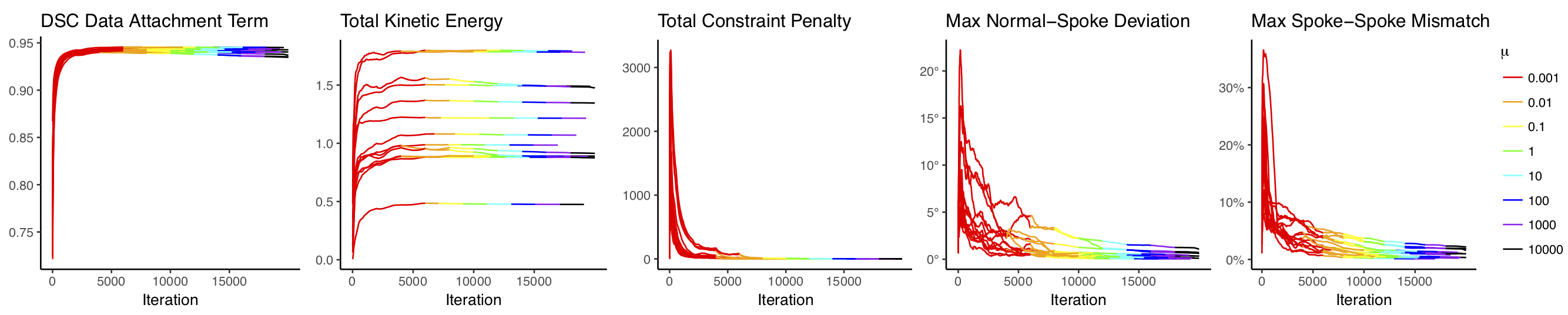}\caption{Data attachment, kinetic energy, and constraints over the course of
AL optimization for the 14 LV wall matching cases. \label{fig:heartplots-1}}
\end{figure}

\subsection{3D Echocardiography Data}

\begin{figure}
\includegraphics[width=1\textwidth]{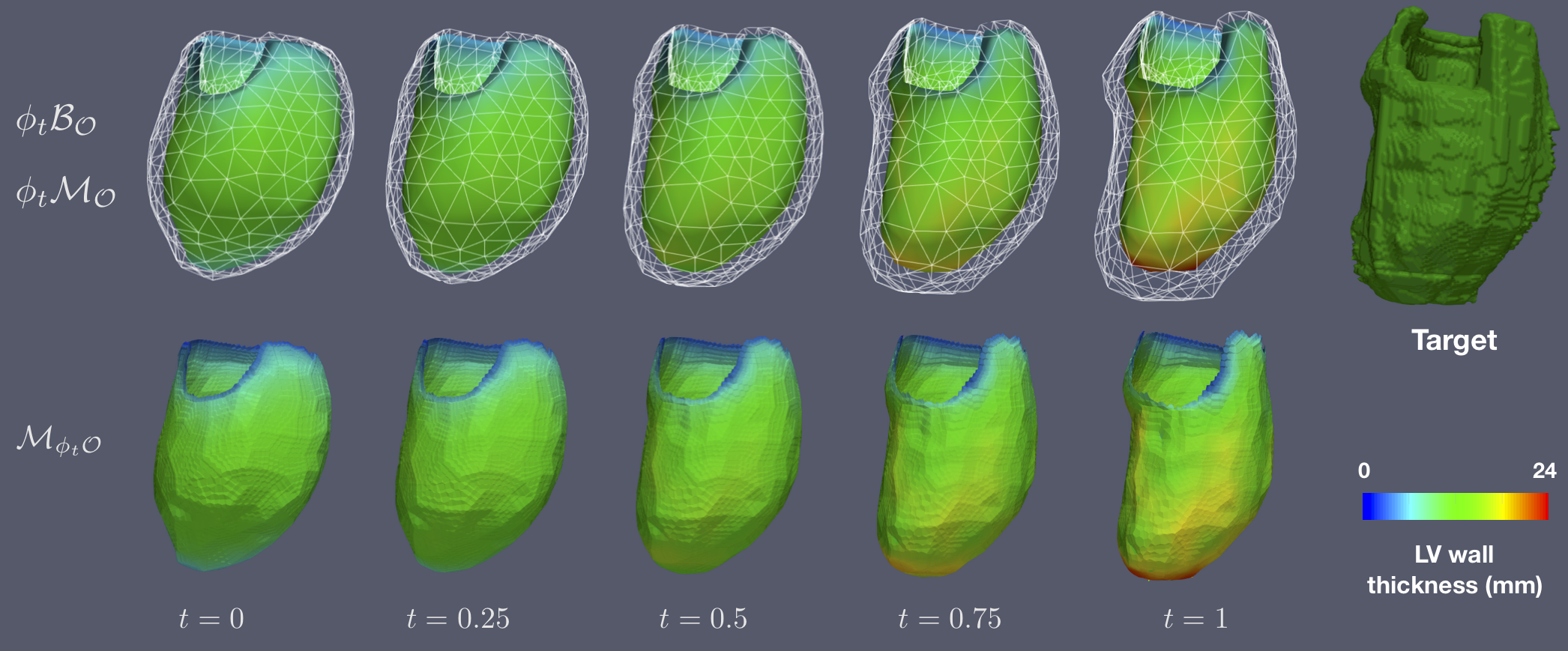}\caption{Example of shape matching in the LV wall dataset. The top row shows
the deforming medial scaffold (colored by thickness = $2\times$ MIB
radius) and the deforming boundary as a white wireframe. The bottom
row shows the Voronoi skeleton of the deforming boundary, again emphasizing
the preservation of medial scaffold structure during deformation.
\label{fig:heartevol-1}}
\end{figure}

To demonstrate the feasibility of diffeomorphic medial modeling on
real-world imaging datasets, we use our approach to model the shape
of the left ventricle (LV) wall in a dataset of 14 transesophageal
3D echocardiograms acquired prior to heart valve surgery. The dataset
includes images from 8 patients who underwent surgery for ischemic
mitral regurgitation (IMR) and 6 patients who did not have surgery.
The LV wall was manually segmented at systole in each case by trained
raters. The first IMR case was used to construct a template medial
model using the same approach as in the toy example. Affine registration
guided by five manually-placed landmarks was performed between the
binary segmentation images of the first IMR case and all other cases.
This was needed to correctly line up the LV outflow tract between
cases. To initialize our medially-constrained diffeomorphometry approach
with the affine registration parameters, we first performed one iteration
of AL optimization with the SSD data attachment term, where the affine-transformed
template landmarks were treated as target locations $q^{\text{trg}}$.
We then performed multiple iterations of AL optimization with the
DSC data attachment term. Fig. \ref{fig:heartplots-1} plots the DSC
data attachment term, kinetic energy, total constraint penalty, maximum
spoke-normal deviation and maximum spoke-spoke mismatch over the course
of AL optimization. Fitting performance was highly consistent across
the 14 cases, with final Dice coefficient ranging between $0.935$
and $0.945$ for all cases, and the maximum normal-spoke deviation
and spoke-spoke mismatch having ranges $0.04^{\circ}-1.08^{\circ}$
and $0.1\%-2.2\%$ respectively. Overall, this demonstrates excellent
ability of medial-preserving diffeomorphic flows to fit real-world
anatomical shapes. Fig. \ref{fig:heartevol-1} shows an example of
the LV template deforming to match an individual case. As in the toy
example, the preservation of the template's medial scaffold is demonstrated
by excellent consistency between the deforming medial scaffold of
the template ($\phi_{t}{\cal M_{O}}$) and the Voronoi skeleton of
the deforming template (${\cal M}_{\phi_{t}{\cal O}}$).

\section{Discussion and Conclusions}

To our knowledge the method presented here is the first attempt to
directly combine medial modeling and diffeomorphometry (an indirect
approach to combining cm-reps and diffeomorphometry was recently proposed
by Hong et al. \cite{hong18}). The current paper demonstrates the
feasibility of shape matching with this approach, but additional work
is needed to extend all the tools of computational anatomy \cite{miller06}
(e.g., computing mean shapes or principal geodesics) to diffeomorphic
medial modeling. A limitation of our approach is its high computational
burden, with thousands of iterations needed for AL optimization to
converge and each iteration having cost quadratic in the number of
landmarks. The heart shape experiments were performed in an hour per
case on a 16-core CPU. An open-source C++ implementation of our method
is at \url{https://github.com/pyushkevich/cmrep}.

\bibliographystyle{plain}
\bibliography{ipmi2019}

\end{document}